\documentclass[%
 reprint,
 superscriptaddress,
 acs
]{revtex4-2}

\usepackage{physics}
\usepackage{graphicx} 
\usepackage{mathtools}
\usepackage[breaklinks=true]{hyperref} 
\usepackage[capitalize]{cleveref}
\usepackage{siunitx}

\newcommand{\HF}{\mathrm{HF}}

\newcommand{\CVS}{\text{CVS}}

\DeclareSIUnit{\au}{{a.u.}}
\DeclareSIUnit\angstrom{\text {Å}}
\DeclareSIUnit\hartree{\text{\ensuremath{E}}_{\mathrm{h}}}

\begin{document}

\title{Coupled cluster simulation of impulsive stimulated x-ray Raman scattering}
 
\author{Alice \surname{Balbi}}   
\thanks{%
These authors contributed equally to this work.
}%
\affiliation{%
Scuola Normale Superiore, Piazza  dei  Cavalieri, 7, I-56126, Pisa, Italy
}%

\author{Andreas S. \surname{Skeidsvoll}}
\thanks{%
These authors contributed equally to this work.
}%
\affiliation{%
Department of Chemistry, Norwegian University of Science and Technology, 7491 Trondheim, Norway
}%

\author{Henrik Koch}
\email{Electronic mail: henrik.koch@sns.it}
\affiliation{%
Scuola Normale Superiore, Piazza  dei  Cavalieri, 7, I-56126, Pisa, Italy
}
\affiliation{%
Department of Chemistry, Norwegian University of Science and Technology, 7491 Trondheim, Norway
}
\date{\today}
%

\begin{abstract}
Time-dependent equation-of-motion coupled cluster (TD-EOM-CC) is used to simulate impulsive stimulated x-ray Raman scattering (ISXRS) of ultrashort laser pulses by neon, carbon monoxide, pyrrole, and p-aminophenol. The TD-EOM-CC equations are expressed in the basis of field-free EOM-CC states, where the calculation of the core-excited states is simplified through the use of the core-valence separation (CVS) approximation. The transfer of electronic population from the ground state to the core- and valence-excited states is calculated for different numbers of included core- and valence-excited states, as well as for electric field pulses with different polarizations and carrier frequencies. The results indicate that Gaussian pulses can transfer significant electronic populations to the valence states through the Raman process. The sensitivity of this population transfer to the model parameters is analyzed. The time-dependent electronic density for p-aminophenol is also showcased, supporting the interpretation that ISXRS involves localized core excitations and can be used to rapidly generate valence wavepackets.
\end{abstract} 

\maketitle
%
%
\section{Introduction} 
The ability to experimentally generate short and intense x-ray laser pulses has been a subject of significant interest in the field of x-ray science. Recent technological advances, specifically the realization of x-ray free electron lasers (XFELs)~\cite{PellegriniXFELs,dev_xfels} and new approaches based on high harmonic generation (HHG)~\cite{mcpherson1987hhg,ferray1988hhg}, have made it possible to generate x-ray laser pulses with high intensities and pulse durations as short as a few hundred and even tens of attoseconds~\cite{shortest_atto}. This progress has enabled the development of new experimental techniques with unprecedented temporal resolution, facilitating the imaging and control of atoms and molecules on the time scale of electronic motion.~\cite{realtimeLeone2014, Baltuska2003, DurisXFEL,ultrafast_xray,atto_li,ultraCalegari,ultraKraus} An important phenomenon in this context is impulsive stimulated x-ray Raman scattering (ISXRS), which is the extension of stimulated x-ray Raman scattering (SXRS) to the impulsive limit, where the duration of the external field interaction is short compared to the time scales of the subsequent evolution of the system.

\begin{figure}
    \centering
    \includegraphics[width=3.375in]{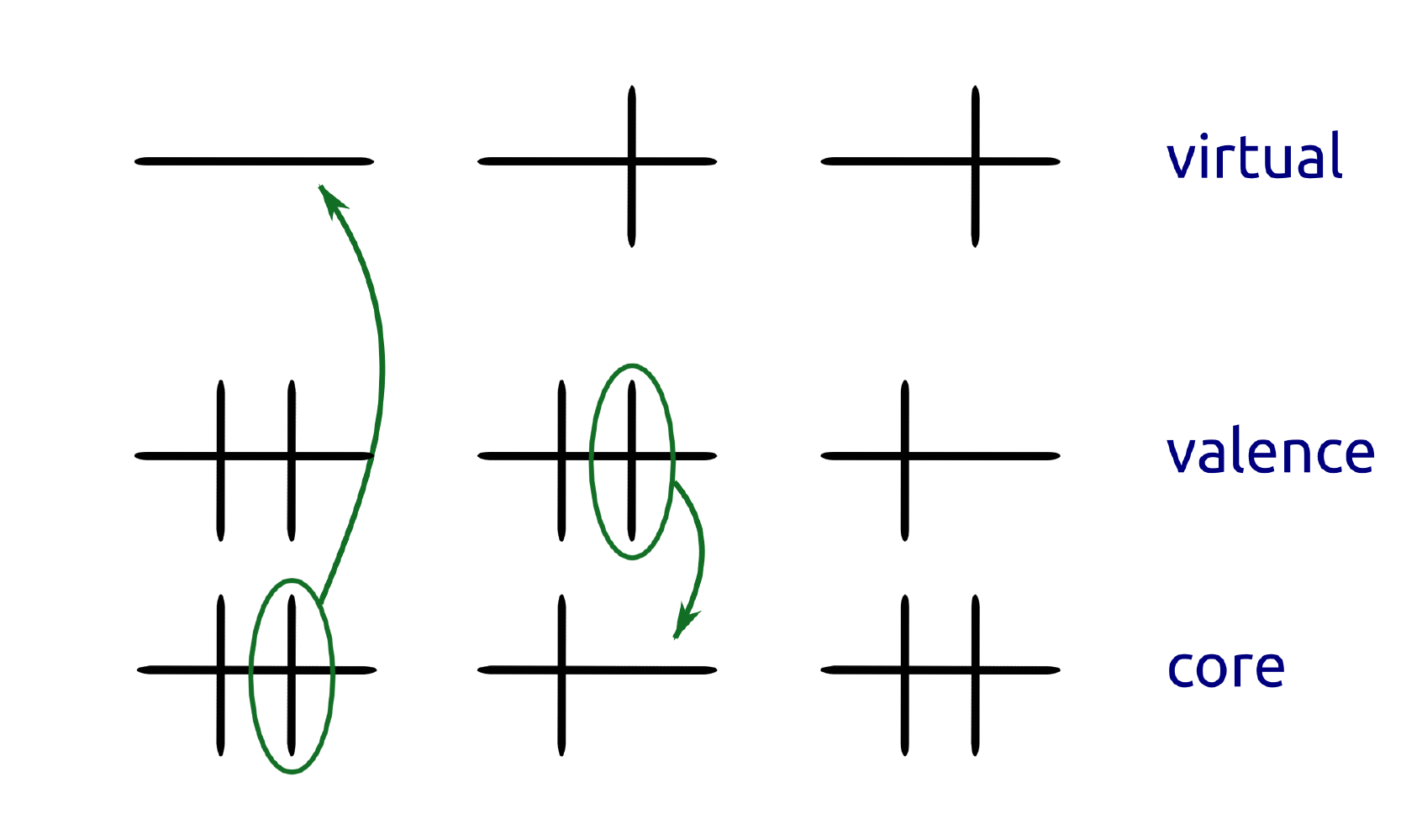}
    \caption{Illustration of the steps in the ISXRS process. Initially, the molecule is in its ground state (left). An external x-ray pulse excites a core electron, leading to a core-excited state (middle). The same pulse can trigger the decay of a valence electron into the core vacancy, leading to a valence-excited state (right).} 
    \label{fig:raman} 
\end{figure}

In general, Raman scattering is a light-matter interaction phenomenon in which photons trigger an excitation of an atomic or molecular system followed by a deexcitation to an energy level different from the initial one. In the context of x-ray Raman scattering, the involved transitions are electronic in character.~\cite{xrayRamanBiggs, xrayRamanTanaka, xrayRamanRohringer, Raman2D, realtimeLeone2014} We focus on the situation in which the electronic excitation in play is a core excitation, which is deexcited to a valence-excited state through the decay of a valence electron into a core vacancy, see \cref{fig:raman}. Core excitations are often localized on a specific atomic site and sensitive to the surrounding electronic environment, making them useful for the local initiation of charge migration. We treat the case where both the excitation and deexcitation are stimulated by an interaction with the same laser pulse.~\cite{xrayRamanWeninger} This is achievable by utilizing a pulse with sufficient bandwidth to encompass the energy differences between the ground state and the core-excited states of interest, as well as between these core-excited states and the final valence-excited states.
The interaction with such pulses is similar to the interactions occurring in the first experimental demonstration of electronic population transfer via ISXRS, which was made for the NO molecule at the Linac Coherent Light Source as recently as in 2020.~\cite{SLAC2020}

The progress in experimental techniques has stimulated the development of methods for modeling electron dynamics based on the time-dependent Schr\"{o}dinger equation. Real-time methods, which involve solving this equation in the real time domain, offer a particularly suitable approach for analyzing ultrafast phenomena.~\cite{li2020real} Among these methods, real-time coupled coupled cluster methods offer high accuracy and computational costs that scale polynomially with the system size. The time evolution is described by differential equations that can be solved using standard numerical integration techniques such as Runge-Kutta methods.

A specific subcategory of real-time coupled cluster methods is the time-dependent coupled cluster (TDCC) methods,~\cite{QD-CC_tdcc3, huber2011tdcc4, pigg2012tdcc5, transient_tdcc6, Park2019tdcc7, kristiansen2020tdcc9, pedersen2021tdcc10, Park2021tdcc11, transient_tdcc6, Pedersen2019symplectic} where the time dependence is parametrized by cluster amplitudes and Lagrange multipliers.~\cite{cc_response1,cc_response2} These methods offer the advantage of size-extensivity at all levels of truncation. Another subcategory, the time-dependent equation-of-motion coupled cluster (TD-EOM-CC) methods,~\cite{nascimento2016linear_tdeomcc, schlegel2007electronic_tdeomcc, luppi2012computation_tdeomcc, Eugene_2017_tdcc1, nascimento2019general_tdeomcc, Koulias_2019_tdcc2} provides less potential for numerical issues compared to TDCC methods,~\cite{skeidsvoll2023comparing} since the time dependence is parametrized by the linear coefficients used in EOM-CC methods and the cluster amplitudes remain fixed at their time-independent ground state values.~\cite{mukherjee1979response_eom1, emrich1981extension_eom2, stanton1993equation_eom3}

In the basis of field-free EOM-CC states, the TD-EOM-CC method requires the predetermination of the excited states that are involved in the studied processes. Computationally, the exterior eigenvalue algorithms usually employed for calculating valence-excited states are inefficient for the calculation of the core-excited states often involved in x-ray interactions. This is because the core-excited states have large eigenvalues, and the states are embedded in an ionization (pseudo-)continuum.~\cite{Norman2018}
A useful scheme for the study of core excitations is the core valence separation (CVS) scheme, which disregards all excitations that do not involve at least one core orbital.~\cite{cederbaum1980firstCVS, Coriani2015} 
This allows for the approximate core-excited states to be calculated as the lowest energy states within the reduced excitation space.

In this article, we use the TD-EOM-CC method together with the CVS approximation to simulate the interaction of neon, carbon monoxide, pyrrole, and p-aminophenol with ultrashort laser pulses, and calculate the populations of the valence-excited states following ISXRS targeting molecular K-edges. The article is organized as follows. In \cref{sec:theory} we briefly outline the theory behind the calculations. We provide details of the performed computations in \cref{sec:comp}, and present and discuss the results in \cref{sec:res}. Conclusions are presented in \cref{conclusion}.  
%
\section{Theory}\label{sec:theory}
%
The time-dependent system is described by the Hamiltonian
\begin{equation}
    H(t) = H^{(0)} + V(t),
\end{equation}
where $H^{(0)}$ is the electronic Hamiltonian of the molecule in the Born-Oppenheimer approximation. We describe the interaction with the external laser field $V(t)$ in the dipole approximation and length gauge,
\begin{equation}
    \label{eq:interactionterm}
    V(t) = -\vb*{d}\vdot\vb*{\mathcal{E}}(t),
\end{equation}
where $\vb*{d}$ is the vector of Cartesian dipole operators, and $\vb*{\mathcal{E}}(t)$ the Cartesian electric field vector.

The eigenstates of the field-free Hamiltonian,
\begin{align}
    \ket{\psi_j} &= \sum_\lambda e^{T}\ket{\lambda}r_{\lambda j} \\
    \bra{\psi_i} &= \sum_\kappa l_{i\kappa}\bra{\kappa}e^{-T}
\end{align}
can be found by first solving the ground state coupled cluster equations
\begin{equation}
    \label{eq:groundstateequations}
    \bra{\mu}e^{-T}H^{(0)}e^{T}\ket{\HF} = 0,
\end{equation}
which determine the cluster amplitudes $t_\mu$ in the cluster operator,
\begin{equation}
    T = \sum_\mu t_\mu \tau_\mu.
\end{equation}
Thereafter, the right and left vectors can be found as eigenvectors of the projected time-independent Schr\"{o}dinger equation,
\begin{align}
    \sum_\lambda\bra{\kappa}e^{-T}H^{(0)}e^{T}\ket{\lambda}r_{\lambda j} &= r_{\kappa j}E_j, \\
    \sum_\kappa l_{i\kappa}\bra{\kappa}e^{-T}H^{(0)}e^{T}\ket{\lambda} &= E_i l_{i \lambda}.
\end{align}
These equations lead to the following eigenvalue problems~\cite{doi:https://doi.org/10.1002/9781119019572.ch13} 
\begin{align}
    \vb{A}\vb{R}_j &= \vb{R}_j\Delta E_j, \\
    \vb{L}_i^T\vb{A} &= \Delta E_i\vb{L}_i^T,
\end{align}
where $A_{\mu\nu} = \bra{\mu}e^{-T}\comm{H^{(0)}}{\tau_\nu}e^{T}\ket{\HF}$, $L_{i\mu}=l_{i\mu}$ and $R_{\nu j}=r_{\nu j}$ for $\mu>0$ and $\nu>0$. The excitation energy $\Delta E_j = E_j - E_0$ is given as the difference between the excited state energy and the ground state energy
\begin{equation}
    E_0 = \bra{\HF}e^{-T}He^T\ket{\HF}.
\end{equation}

The TD-EOM-CC ket and bra states can be expanded in the field-free EOM-CC kets and bras, $\ket{\Psi(t)}=\sum_j \ket{\psi_j}c_j(t)$ and $\bra*{\widetilde{\Psi}(t)}=\sum_i b_i(t)\bra*{\widetilde{\psi}_i}$. This gives the TD-EOM-CC equations~\cite{PhysRevA.105.023103}
\begin{align}
    \label{eq:rightevolution}
    i \dv{c_i(t)}{t} &= \sum_j H_{ij}(t)c_j(t), \\
    \label{eq:leftevolution}
    -i \dv{b_j(t)}{t} &= \sum_i b_i(t)H_{ij}(t),
\end{align}
where $H_{ij}(t) = \bra*{\widetilde{\psi}_i}H(t)\ket*{\psi_j} = \delta_{ij}E_j + \bra*{\widetilde{\psi}_i}V(t)\ket*{\psi_j}$. The time-dependent population of EOM-CC state $i$ in the TD-EOM-CC superposition state can be found as the product of the projections onto the ket and bra of the EOM-CC state,
\begin{equation}
    \begin{split}
        P_i(t) &= \bra*{\widetilde{\Psi}(t)}\ket{\psi_i}\bra*{\widetilde{\psi}_i}\ket{\Psi(t)} \\
        &= b_i(t)c_i(t).
    \end{split}
\end{equation}

The eigenvalues of core-excited states are interior to the spectrum of the molecular Hamiltonian, and often hard to reach using exterior eigenvalue methods like Davidson or Lanczos algorithms. The core-valence separation (CVS) approximation~\cite{cederbaum1980firstCVS, BarthCVS} simplifies the calculation of these states by removing the valence-core and core-valence blocks of the Hamiltonian and has become a vital tool for the calculation of NEXAFS spectra.~\cite{Norman2018}
Let $I$ denote the set indexing the core orbitals. We invoke the CVS approximation through a projector $\mathcal{P}_I^\CVS$ that removes all vector elements that do not reference excitations from at least one core orbital, in each eigensolver iteration.~\cite{Coriani2015}
For the coupled cluster singles and doubles (CCSD) truncation level, this can be expressed in compact form as
\begin{align}
    \mathcal{P}_I^\CVS r_i^a = l_i^a \mathcal{P}_I^\CVS &= 0 \qc i \notin I \\
    \mathcal{P}_I^\CVS r_{ij}^{ab} = l_{ij}^{ab}\mathcal{P}_I^\CVS &= 0 \qc i \notin I \wedge j \notin I. 
\end{align} 
This projection is effectively setting all elements of the valence-valence block of the full-space elementary basis EOM-CC Jacobian matrix $\vb{A}$ to zero, giving the CVS approximated Jacobian matrix, $\vb{A}^{\CVS}$.
The core-excited EOM-CC states obtained in the CVS approximation can have a non-zero overlap with EOM-CC states obtained without invoking this approximation. The CVS states are in general also not eigenstates of the full field-free Jacobian, and can lead to TD-EOM-CC populations that are non-stationary, complicating the interpretation of the TD-EOM-CC state. To ensure that the populations are stationary, we diagonalize the Jacobian $\vb{A}$ in the basis of all the CVS and non-CVS (valence) states by first constructing the Jacobian and overlap matrices
\begin{equation}
    A_{ij} = \vb*{L}_i\vb{A}\vb*{R}_j \qc S_{ij} = \vb*{L}_i\vb*{R}_j.
\end{equation}
respectively in the reduced space. Assuming linear independence of the vectors in the basis, the solution of the generalized eigenvalue problem defined by $\vb{A}$ and $\vb{S}$ gives a new set of right and left eigenvectors of $\vb{A}$, which preserve populations when there is no interaction with the external field.
%
\section{Computational details}\label{sec:comp}
%
The electric field in \cref{eq:interactionterm} is represented as
\begin{equation}
    \vb*{\mathcal{E}}(t) = \vb*{\mathcal{E}}_0\cos(\omega_{0}(t-t_0) + \phi)f(t),
\end{equation}
where $\vb*{\mathcal{E}}_0$ is the peak electric field of the pulse in its polarization direction, $\omega_0$ the carrier frequency and  $t_0$ the central time of the pulse, and $\phi$ is the carrier-envelope phase. The envelope function $f(t)$ is chosen to have the Gaussian shape
\begin{equation}
    \label{eq:envelope}
    f(t) =
    \begin{cases}
        e^{-(t - t_0)^{2}/(2\sigma^{2})}, & -c \le t \le c, \\
        0, & \text{otherwise},
    \end{cases}
\end{equation}
where the RMS width is selected as $\sigma=0.5$ 
and the envelope truncated at $c = 8 \sigma $.
In all calculations, we use the carrier-envelope phase $\phi=0$ and the peak electric field strength $\abs{\vb*{\mathcal{E}}_0}$ of \SI{10}{\au}, which corresponds to the maximum intensity of \SI{7.019e18}{\watt\per\square\centi\meter}, calculated from the intensity relation $S_{0} = \abs{\vb*{\mathcal{E}}_{0}}^{2}/Z_{0}$ where $Z_{0}$ is the impedance of free space.  

All simulations are performed using a development version of the eT program~\cite{eTprogram} containing the TD-EOM-CC implementation described in Ref.~\cite{PhysRevA.105.023103}. The Runge-Kutta method known as RK4 is used to integrate \cref{eq:rightevolution} and \cref{eq:leftevolution}, with time steps of \SI{0.001}{\au} for neon, carbon monoxide, and p-aminophenol and \SI{0.0001}{\au} for pyrrole.
%
\section{Results and discussion}\label{sec:res}
%
\subsection{Neon}
\begin{figure*} 
    \centering
    \includegraphics[width=7in]{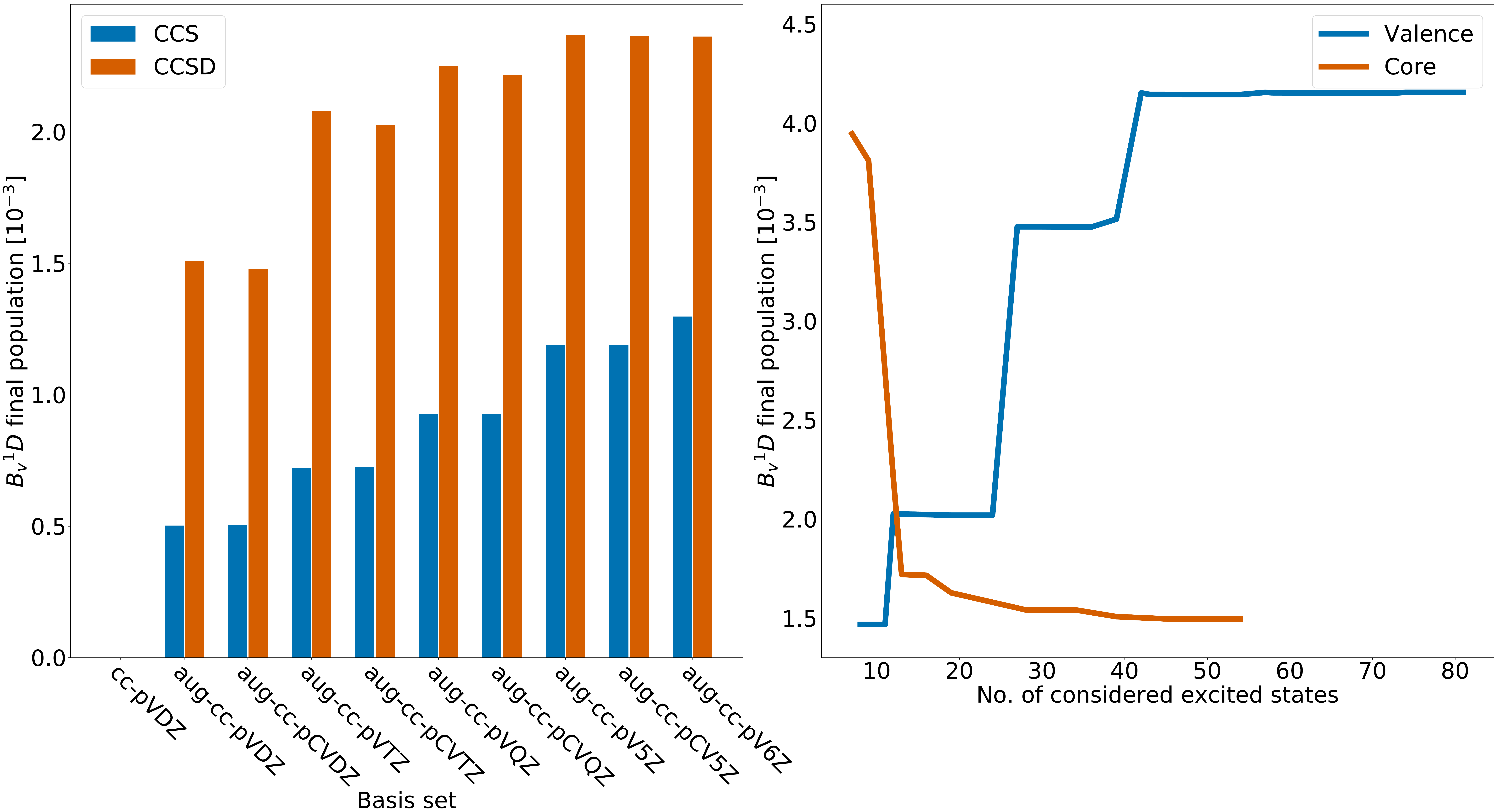}
    \caption{The left panel shows the final population of the $B_v\prescript{1}{}D$ states of neon for different choices of level of theory and basis set. The blue line in the right panel shows the final population of the same $B_v\prescript{1}{}D$ states for different numbers of valence-excited states included in the simulation and the number of core-excited states fixed at 4, calculated with CCSD/aug-cc-pCVTZ. The red line in the right panel shows the final population of the same $B_v\prescript{1}{}D$ states for different numbers of core-excited states included in the simulation and the number of valence-excited states fixed at 79, calculated with CCSD/aug-cc-pCVTZ.} 
    \label{fig:neon} 
\end{figure*}
In the following, the convergence properties of the final Raman-induced populations are investigated for the neon atom. This system is used for benchmarking purposes, as its small size allows for the use of larger basis sets. We focus on the convergence of the final population of the $B_v\prescript{1}{}D$ valence-excited state, the lowest valence-excited state with a significant final population. 

We first study the basis set convergence with respect to the cardinal number X of Dunning basis sets for CCS and CCSD levels of theory. The employed basis sets are cc-pVDZ, aug-cc-pVXZ (with X=D,...,6) and aug-cc-pCVXZ (with X=D,...,5).  
As the carrier frequency $\omega_0$ of the electric field, we choose the average of two frequencies. The first frequency corresponds to the transition between the ground state $X {}^1S$ and the $B_c {}^1P$ core-excited state. The second frequency corresponds to the transition between the $B_c {}^1P$ core-excited state and the $B_v {}^1 D$ valence-excited state.
The $B_c {}^1P$ and $B_v {}^1 D$ states are chosen as they are, respectively, the lowest core-excited and valence excited states that get significantly populated in the Raman process, except for the cc-pVDZ basis set, where the order of $A_c {}^1S$ and $B_c {}^1P$ energy levels is inverted. In these calculations, we include 4 core-excited states and 12 valence-excited states. The frequencies used for the different basis sets and levels of theory are given in the Supporting Information. 


%
%
%
From \cref{fig:neon}, we can observe how the final populations calculated with CCS and CCSD are considerably different, implying that CCS is not accurate enough to provide an adequate description of the system. The addition of functions for describing core correlation (aug-cc-pCVXZ) leads to slightly lower final populations compared to the corresponding basis sets without these functions (aug-cc-pVXZ). For CCSD, the results for 5Z and 6Z are very similar, implying that basis-set convergence is reached for 5Z.
Continuing, the convergence of the final population of the $B_v\prescript{1}{}D$ state is explored with respect to the number of valence- and core-excited states included in the calculation. The total of the probabilites of all degeneracies of a state is calculated, such that for instance the probabilities for the five degenerate states of $D$ type are added together. We perform the calculations using the CCSD truncation level and the aug-cc-pCVTZ basis set.
The right panel of \cref{fig:neon} exhibits the convergence of the final population of the $B_v\prescript{1}{}D$ states with respect to the number of valence-excited states included in the simulation, with the number of core-excited states fixed at 4. The results indicate that more than 40 valence-excited states are needed for convergence.
An analogous procedure is performed, this time keeping the number of valence-excited states fixed while varying the number of core-excited states. In the right panel of \cref{fig:neon}, we can see how the final population of $B_v\prescript{1}{}D$ starts to converge after around 15 core-excited states are included in the calculation.
\subsection{CO}
\begin{figure*}
    \centering
    \includegraphics[width=7in]{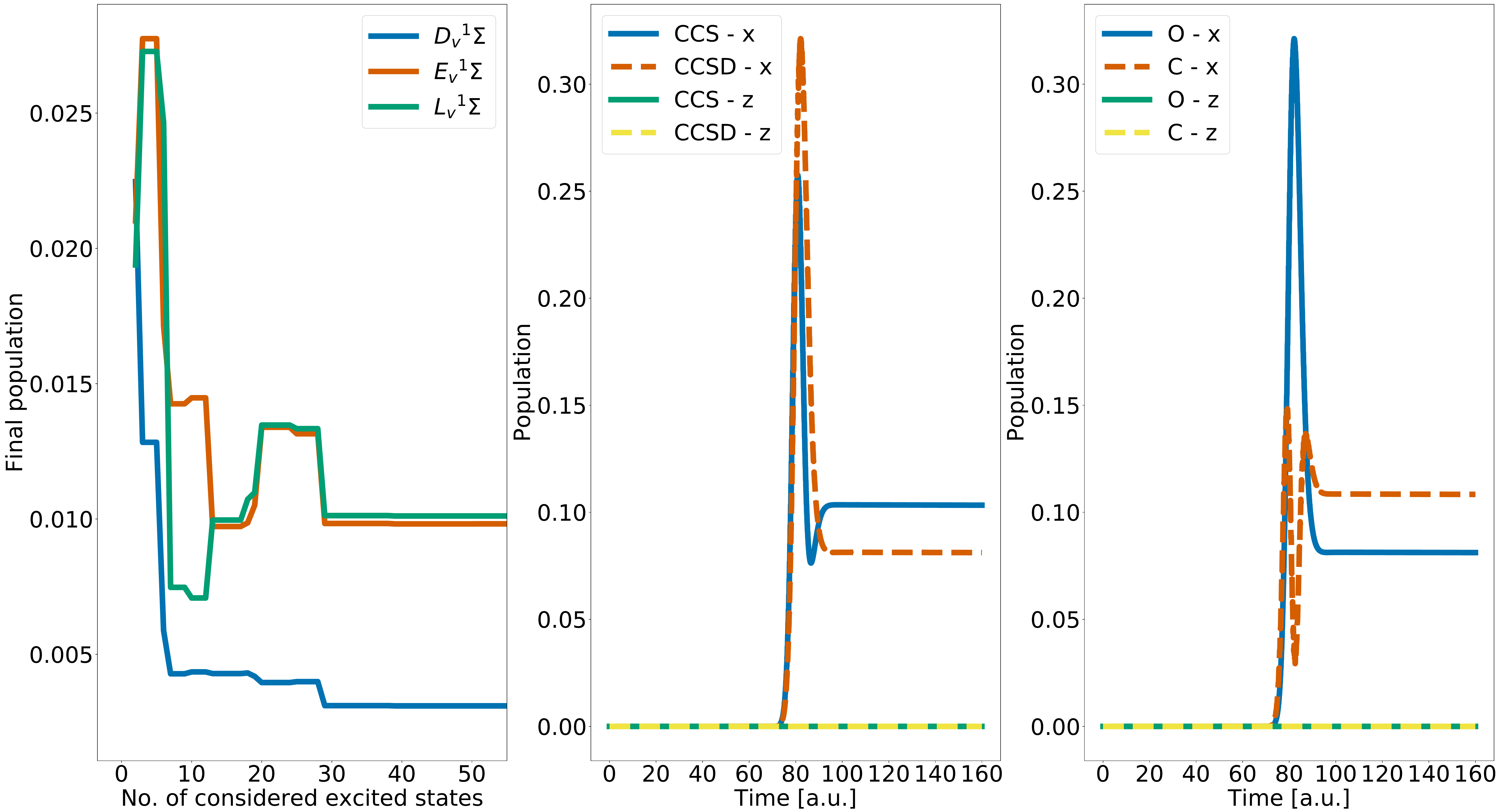} 
    \caption{The left panel shows the final population of the $D_v{}^1\Sigma$, $E_v{}^1\Sigma$, and $L_v{}^1\Sigma$ valence-excited state of carbon monoxide for different numbers of core-excited states included in the calculation, with the number of valence states fixed at 20 and the external electric field polarization in the positive $z$-direction. The central panel shows the time-dependent population of the third valence-excited state of carbon monoxide, calculated with the aug-cc-pCVTZ basis set and different levels of theory and electric field polarizations. The right panel shows the time-dependent population of the third valence-excited state for external electric fields tuned to different K-edges and with different polarizations, calculated with CCSD/aug-cc-pCVTZ.}
    \label{fig:co}
\end{figure*}
We continue by simulating ISXRS for the carbon monoxide molecule, which is linear and belongs to the $C_{\infty v}$ symmetry point group. Since the system is not centrally symmetric, results can differ depending on the polarization of the electric field. Theoretical and experimental studies of the core-excitation spectroscopy and ISXRS of this molecule have previously been carried out.~\cite{coreexcCO}
In our simulations, the distance between the two nuclei is fixed at $\SI{1.128}{\angstrom}$, corresponding to the equilibrium bond length in the NIST database.~\cite{nist_diatomic}. The internuclear axis of the molecule is aligned along the $z$-axis and the carbon atom is placed at the origin of the coordinate system while the oxygen atom is placed at $\SI{1.128}{\angstrom}$ along the $z$-axis. The carrier frequency of the external electric field is again chosen as the average between two frequencies. The first is the transition frequency between the ground state and the first core-excited state, which is the lowest-energy core-excited state that gets significantly populated during the Raman process. The second is the frequency of transition between this core-excited state and the third valence-excited state, which is the lowest valence-excited state that gets significantly populated. For CCS/aug-cc-pCVTZ, the frequency is \SI{20.029089}{\hartree}, while for CCSD/aug-cc-pCVTZ it is \SI{19.504022}{\hartree}, corresponding to the O K-edge.

To investigate transitions at the C K-edge, we choose the lowest-energy molecular orbital localized on the carbon atom as the molecular orbital used in the CVS approximation. The carrier frequency of the electric field is again chosen as average of the transition frequencies between the ground state and the lowest core-excited state that is significantly populated, and that between that core-excited state and the lowest valence-excited state that is significantly populated, resulting in a carrier frequency of \SI{10.402530}{\hartree} 

In the carbon monoxide system, linearly polarized electric fields can be decomposed into two components: the polarization component parallel to the internuclear axis (along the $z$-axis) and the polarization component perpendicular to it (any direction in the $xy$-plane). As for neon, the convergence of the final population of certain valence-excited states is assessed with respect to the number of included core-excited states. The results are shown for the $D_v{}^1\Sigma$, $E_v{}^1\Sigma$ and $L_v{}^1\Sigma$ valence-excited states in the left panel of \cref{fig:co}, demonstrating that convergence is attained by increasing the number of considered core-excited states. About 30 core-excited states are needed for convergence when the number of valence-excited states is fixed at 20. In the central panel of the figure, we can see how the time-dependent population of the third valence-excited state depends on the polarization of the electric field and level of theory, and also how the population is constant after the interaction with the field. The final population is exactly zero when the polarization is along the $z$-axis, as expected from the symmetry of the molecule and field. In the right panel, we can see how the time-dependent population of the third valence-excited state differs when the carrier frequency of the electric field is tuned to the K-edge of different elements (C or O). For the different tunings, the third valence-excited state is reached through different transition pathways, involving other transition frequencies and transition moments. As for the results in the central panel, the population is exactly zero when the electric field is polarized along the $z$-axis, irrespective of the chosen frequency, for symmetry reasons.  The populations are also constant after the interaction with the field

\subsection{Pyrrole}
\begin{figure*}
    \centering
    \includegraphics[width=7in]{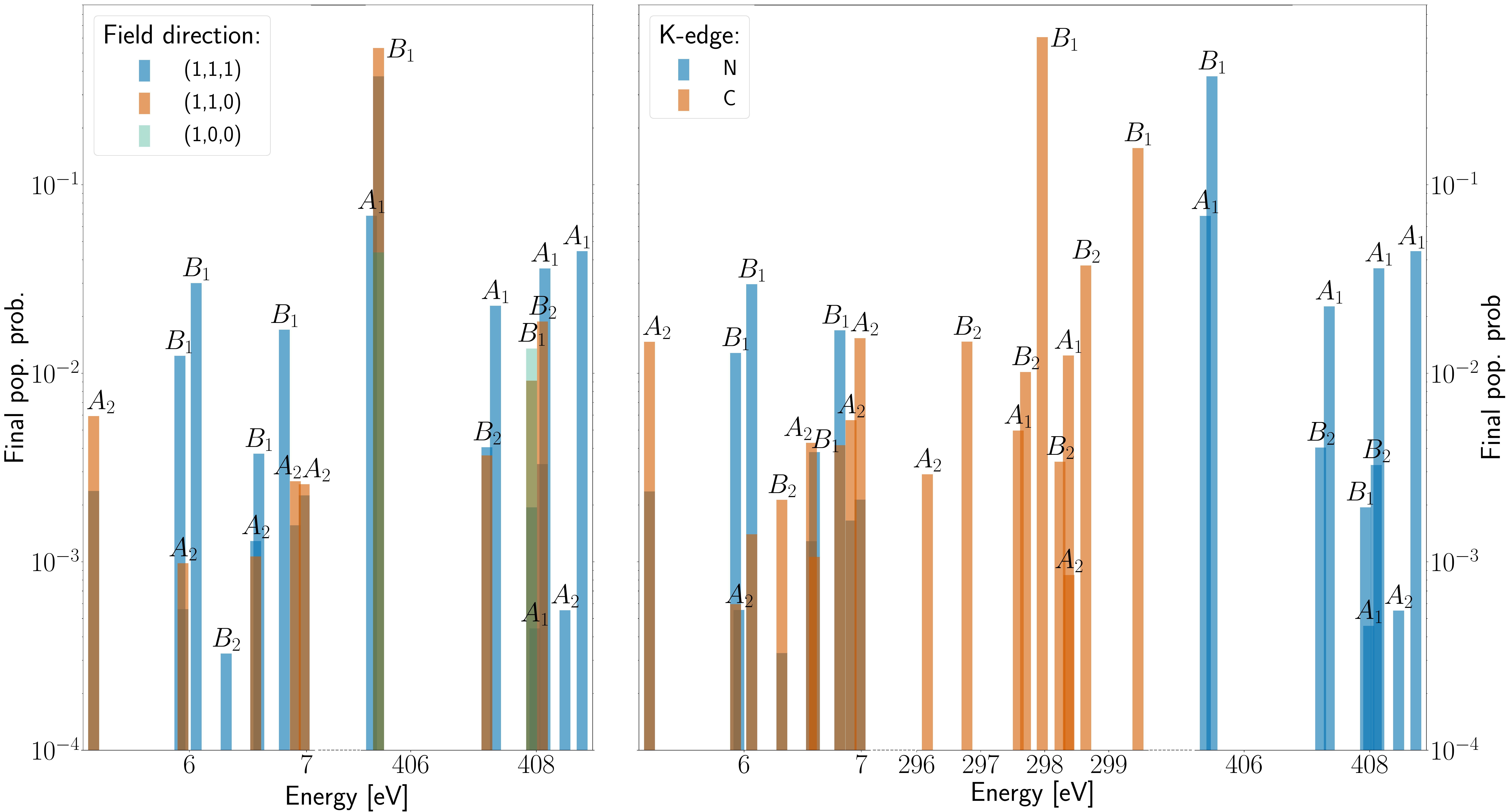} 
    \caption{The left panel displays the final populations of various excited states of pyrrole following ISXRS with different electric field polarizations, computed at the CCSD level of theory and with the aug-cc-pCVDZ basis set for the nitrogen atom and the aug-cc-pVDZ basis set for the other atoms. The right panel displays the final populations of different excited states of pyrrole for electric fields tuned to different K-edges, computed at the CCSD level of theory and with the aug-cc-pCVDZ basis set for the atom with the targeted K-edge shown in the inset and aug-cc-pVDZ basis set for the remaining atoms.} 
    \label{fig:pyrrole} 
\end{figure*}
We further increase the complexity of the modeled system by considering pyrrole, which belongs to the $C_{2v}$ symmetry point group. The geometry of the molecule is obtained from the NIST database.~\cite{nist_diatomic}, for which the molecule lies in the $yz$-plane and the symmetry axis is along the $z$-axis. The Supporting Information provides the geometry of the system, along with a figure that shows its orientation relative to the Cartesian coordinate axes. The final populations after the Raman process are assessed for the electric field polarization vector set equal to $(1,0,0)$ $(1,1,0)$ and $(1,1,1)$ in the chosen coordinate system. The Raman process involving the N K-edge is studied by performing calculations at the CCSD level of theory with aug-cc-pCVDZ for the nitrogen atom and aug-cc-pVDZ for the other atoms. The carrier frequency of the external electric field is chosen as the frequency of transition from the ground state to the most populated core-excited state, which is \SI{14.901363}{\hartree}. The Raman process involving the C K-edge is studied by performing calculations at the CCSD level of theory with aug-cc-pCVDZ for the carbon atoms and aug-cc-pVDZ for the other atoms. The core-excited states are calculated by using the CVS approximation restricted to the molecular orbital with the second-lowest energy. The carrier frequency of the external electric field is set to \SI{10.949885}{\hartree}, which is the transition frequency from the ground state to the fifth core-excited state, the lowest-energy core-excited state that is the most populated.

In the left panel of \cref{fig:pyrrole}, we can see that new valence-excited states are populated as the polarization of the external electric field changes from $(1,0,0)$, to $(1,1,0)$, and to $(1,1,1)$. In particular, when the electric field is only polarized along the $x$-axis, there are no excitation to the valence-excited states. When the electric field has components along all three axes, all considered valence-excited states have a nonzero final population. An intermediate situation occurs when the electric field has components along both the $x$- and $y$-axes but not along the $z$-axis. This is since the different polarizations of the external electric field has components in different numbers of irreducible representation, enabling transitions to electronic states belonging to different irreducible representations.
In the right panel of \cref{fig:pyrrole}, we can see how the final population of valence-excited states differs when the carrier frequency of the electric field is tuned to the $N$ K-edge and $C$ K-edge, calculated using the CVS approximation with the lowest- and next-to-lowest-energy molecular orbitals, respectively. In both cases, the polarization vector of the field is set to $(1,1,1)$. The valence-excited states that become populated are the same for the two K-edge frequencies, while the populations of the states are different.
\subsection{\textit{p}-aminophenol}

\begin{figure*}
    \centering
    \includegraphics[width=7in]{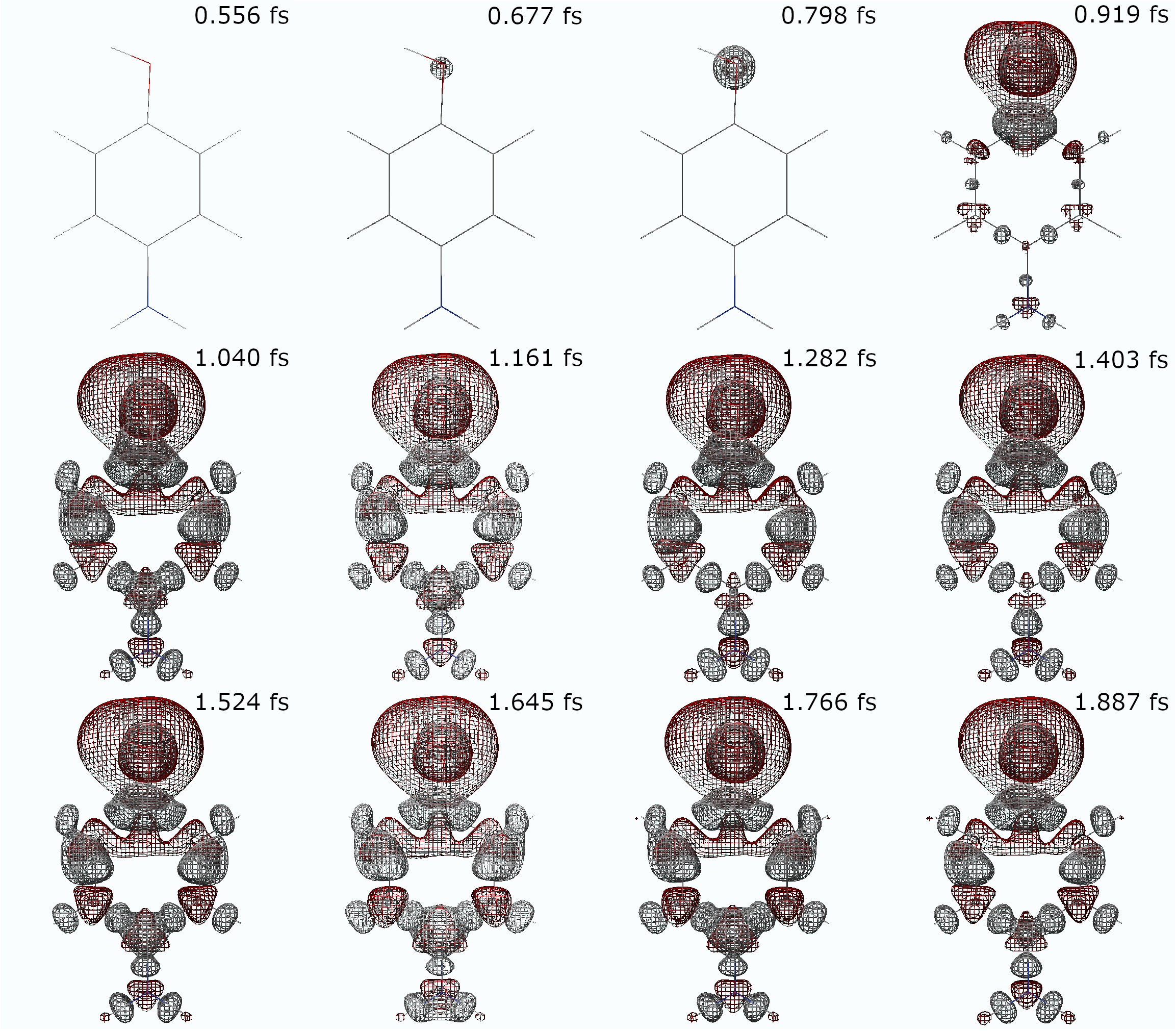}
    \caption{Positive (gray) and negative (red) electronic isodensity surfaces of the time-dependent density after subtracting the ground state density of \textit{p}-aminophenol, at the times specified at the top right corner of each subfigure. The structure of the \textit{p}-aminophenol molecule is also shown in each subfigure.} 
    \label{fig:pap_dens_timefs} 
\end{figure*}
Finally, we consider the planar \textit{p}-aminophenol molecule. The molecule belongs to the $C_s$ symmetry point group, which only contains the mirror plane and the identity as symmetry elements. This molecule is chosen in order to investigate if charge migration between the functional groups located at the opposite side of the aromatic ring can be observed, as the electronic charge can easily travel along the aromatic electron cloud.~\cite{chargetransfer}

Compared to the systems analyzed previously, which offer only limited potential for charge migration due to their small sizes, the \textit{p}-aminophenol molecule is a larger system containing two strongly electron donor substituents (amino and hydroxyl) on a benzene ring.~\cite{brinck1997ortopara}. We can thus expect a localized excitation to be followed by long-range charge migration.

The geometry of \textit{p}-aminophenol is calculated at the B3LYP/aug-cc-pVDZ level of theory, and the molecule is placed in the $xy$-plane. The Supporting Information includes the geometry and a figure that illustrates the orientation of the molecule relative to the Cartesian coordinate axes. For the subsequent calculations, aug-cc-pCVDZ is used for the oxygen atom and aug-cc-pVDZ for all other atoms. The carrier frequency is chosen as \SI{19.883479}{\hartree}, which corresponds to the frequency of transition from the ground state to the fourth core-excited state, which is the most populated state among the two lowest-energy core-excited states that have a non-zero population after the Raman process.

In \cref{fig:pap_dens_timefs}, the charge migration is illustrated through isodensity surfaces of the time-dependent density after subtracting the ground state density, calculated at different points in time. After the interaction with the external electromagnetic pulse, we can observe how the core excitation of the oxygen atom is reflected in a positive charge arising around that nucleus, enclosed in a negatively charged region at a bigger distance from the oxygen nucleus. This is followed by an alternating pattern of regions with increased or decreased electronic charge throughout the entire benzene ring up to the nitrogen atom of the amino group. In particular, the atoms of the ring gain some negative charge while the bonds become more positively charged, and the bonds are thus expected to be weakened. Finally, we can observe how the nitrogen atom becomes negatively charged. As predicted, we observe a localized excitation at the hydroxyl substituent following oxygen K-edge excitation, followed by long-range charge migration, in accordance with what one could expect from a superposition of valence-excited states generated by ISXRS.

In the supplemental material we have included a movie that shows the temporal evolution of the electronic density depicted through isodensity surfaces of the time-dependent density difference, illustrating how the density oscillates after the interaction with the external electric field. The generation of electronic wavepackets with external laser pulses is interesting from an experimental point of view, as it represents the first step of controlling chemical reactions with laser pulses.
\section{Conclusion}\label{conclusion}
In this work, a time-dependent equation-of-motion coupled cluster model of ISXRS has been presented. First, we assessed the convergence of the final population of neon valence states with respect to different calculation parameters: the level of coupled cluster theory, the choice of basis set, and choices of the total number of valence- and core-excited states. We observed how the adequate description of the system required a proper representation of correlation and a sufficiently flexible basis set, since the CCS level of theory and basis sets without augmentation performed poorly. We also demonstrated that convergence of the population of a valence-excited state of neon was achieved when increasing the number of valence- and core-excited states for the given level of theory and basis set. Subsequently, the final populations of carbon monoxide states were assessed with respect to the number of included core-excited states. The results showed convergence for several valence-excited states for the given level of theory and basis set.

Furthermore, we demonstrated that the final populations of states of both carbon monoxide and pyrrole are significantly affected by the polarization of the external electric field, as symmetry can enable and forbid the transition to some of the excited states within the bandwidth of the pulse. We also assessed how the results were affected by tuning the external electric field to the K-edge of the different atoms, where the frequencies were calculated with the CVS approximation targeting the core molecular orbitals of the atoms. We observed how a different choice of K-edge led to changes in final populations as the final states were reached through different transition pathways.

After investigating ISXRS by neon, carbon monoxide, and pyrrole, we studied the time evolution of the electronic density of \textit{p}-aminophenol. The ground-state density was subtracted from the time-dependent density, and the density difference was visualized through isodensity surfaces in real space. We observed the rapid formation of a valence wavepacket and subsequent charge migration in the molecule.
Simulations of field-induced charge migration in molecular systems can be used to predict how chemical reactions can be controlled by external electric fields, which we believe will be a subject of further interest in the near future.
%
\begin{acknowledgments}
%
We acknowledge Gioia Marazzini for helpful discussions. We acknowledge the financial support from The Research Council of Norway through FRINATEK Project No. 275506. Computing resources provided by Sigma2---the National Infrastructure for High Performance Computing and Data Storage in Norway (Project No. NN2962k) and the Center for High Performance Computing (CHPC) at SNS are also acknowledged.
\end{acknowledgments}
\bibliography{bib}

\end{document}